\documentstyle[prb,aps]{revtex}
\begin{document}
\draft
\title{Electron Dynamics in Nd$_{1.85}$Ce$_{.15}$CuO$_{4+\delta}$:
Evidence for the Pseudogap State and Unconventional c-axis Response}
 \author{E.J.~Singley, D.N.~Basov}
 \address{Department of Physics University of California San Diego,
La Jolla, CA 92093-0319}
\author{K.~Kurahashi$^*$, T.~Uefuji, K.~Yamada}
\address{Institute for Chemical Research, Kyoto University, Uji 611-0011, Japan}

\wideabs{
\maketitle

\begin{abstract}

Infrared reflectance measurements were made with light polarized
along the a- and c-axis of both superconducting and
antiferromagnetic phases of electron doped
Nd$_{1.85}$Ce$_{.15}$CuO$_{4+\delta}$. The results are compared to
characteristic features of the electromagnetic response in hole
doped cuprates. Within the CuO$_2$ planes the frequency dependent
scattering rate, 1/$\tau(\omega)$, is depressed below $\sim$ 650
cm$^{-1}$; this behavior is a hallmark of the pseudogap state.
While in several hole doped compounds the energy scales associated
with the pseudogap and superconducting states are quite close, we
are able to show that in Nd$_{1.85}$Ce$_{.15}$CuO$_{4+\delta}$ the
two scales differ by more than one order of magnitude. Another
feature of the in-plane charge response is a peak in the real part
of the conductivity, $\sigma_1(\omega)$, at 50-110 cm$^{-1}$ which
is in sharp contrast with the Drude-like response where
$\sigma_1(\omega)$ is centered at $\omega=0$. This latter effect
is similar to what is found in disordered hole doped cuprates and
is discussed in the context of carrier localization. Examination
of the c-axis conductivity gives evidence for an anomalously broad
frequency range from which the interlayer superfluid is
accumulated. Compelling evidence for the pseudogap state as well
as other characteristics of the charge dynamics in
Nd$_{1.85}$Ce$_{.15}$CuO$_{4+\delta}$ signal global similarities
of the cuprate phase diagram with respect to electron and hole
doping.\end{abstract}}

\narrowtext

\section{Introduction}

The family of high temperature superconductors A$_{2-x}$Ce$_x$CuO$_4$, where
A is a rare earth ion (Nd, Pr, Sm, Eu), has historically been considered an
exception among copper oxides. Like all cuprates the basic
building blocks of the structure are the CuO$_2$ layers. An important
difference is that in the superconducting phases of A$_{2-x}$Ce$_x$CuO$_4$ lack
the apical oxygen above the in-plane copper atom found in most hole doped
cuprates. The charge carriers in A$_{2-x}$Ce$_x$CuO$_4$ are electrons rather
than holes as in all other cuprate families.\cite{electron,twoband} These
materials have a relatively low T$_c$ and early microwave measurements
suggested the order parameter was $s$-wave\cite{Wu93} in contrast with the
$d$-wave symmetry established for hole doped compounds. However, more recent
microwave measurements \cite{Kokales00,Prozorov00}, along with
photoemission\cite{Armitage00}, and phase sensitive experiments\cite{dwave}
indicate that the order parameter in Nd$_{1.85}$Ce$_{.15}$CuO$_{4+\delta}$
(NCCO) and Pr$_{1.85}$Ce$_{.15}$CuO$_{4+\delta}$ may in fact be $d$-wave. The
most comprehensive infrared work on superconducting NCCO by Homes et
al.\cite{Homes97} found weak electron-phonon coupling below T$_c$,
suggesting that NCCO, like other cuprates, is not a phonon-mediated
superconductor. However, in contrast to other cuprates the in-plane
superfluid was found to be anomalously large for a low T$_c$ material,
causing NCCO to deviate significantly from universal Uemera plot\cite{Uemera89}.
These results leave the nature of the relation between electron and hole
doped cuprates ambiguous.

Previous doping dependent infrared studies have found that the evolution of
spectral weight from the insulting parent compounds through the
superconducting phases is similar in both electron and hole doped cuprates.
\cite{Cooper90,Uchida91,Lupi} Doping first moves spectral weight from above
the charge transfer gap to the mid-infrared, and then to the Drude peak at
$\omega=0$. This implies that whether doped with electrons or holes the
Mott-Hubbard insulator shows the same gross features in the electronic
conductivity. Detailed studies on a variety of hole doped cuprates reveal a
partial gap (pseudogap) in the spectrum of the low energy
excitations.\cite{Timusk99} The pseudogap state is recognized as one of the
key characteristics of the (hole doped) cuprates and is believed to be
intimately connected to the origin of high-$T_c$
superconductivity.\cite{Buchanan01} So far, there have been no reports for a
similar pseudogap region on the electron doped side of the phase
diagram.\cite{hpgap} With this in mind it is critical to determine whether
there are  principal differences in the fundamental interactions defining
carrier dynamics and superconductivity in electron and hole doped Mott-Hubbard
insulators. In short the question "Is NCCO a high temperature
superconductor?"\cite{Stadlober95} needs to be re-visited.

In this work we have determined the optical constants of
Nd$_{1.85}$Ce$_{.15}$CuO$_{4+\delta}$ for both the as grown
antiferromagnetic (AF) phase and oxygen reduced superconducting (SC)
samples. Special attention is paid to the low energy ($\omega<$ 100 meV)
physics. Through an analysis of the in-plane scattering rate,
1/$\tau(\omega)$, we find compelling evidence for a pseudogap in the electron
doped
materials. The doping and temperature dependence of the pseudogap is shown
to mirror the behavior found in hole doped cuprates. Another important
result is concerned with the non-monotonic behavior of $\sigma_1(\omega)$ in
the far-infrared which is similar to that of disordered hole doped cuprates
and suggests charge carrier localization. We have also examined the
interlayer c-axis conductivity of NCCO. A sum rule analysis demonstrates
that nearly all of the interlayer superfluid is accumulated from an energy
scale in excess of 8$\Delta$, where $\Delta$ is the superconducting gap.

This paper is organized in the following manner. Section II gives a brief
overview of the experimental procedures.
The raw R($\omega$) data and the Kramers-Kronig generated
$\sigma_1(\omega)$ for both the a- and c-axis of the SC and AF samples are
presented in Section III. Section IV follows with a discussion of the key
results.  In sub-section IV-A we establish
the existence of a pseudogap through the analysis of the in-plane
scattering rate and discuss the implications of this result. Next, evidence for
localization in the cuprates is presented in sub-section IV-B, and we elaborate
on the impact of localization on both the DC and AC transport properties. Finally,
sub-section IV-C discusses the energy scale related to the interlayer
superfluid response.
We conclude by summarizing our results in section V.

\section{Experimental Procedure}

Single crystals of NCCO were
prepared by the traveling-solvent-floating-zone method.\cite{Kurahashi99}
The as-grown crystals are not superconducting, but show antiferromagnetic
order with T$_N$ = 125 K - 160 K.\cite{Matsuda92} Superconductivity is
achieved by deoxygenating the crystals. This process removes a small amount
of apical oxygen atoms which are absent in the ideal T' structure.
\cite{Jandl95}

The near normal reflectance was measured in polarized light from the
far-infrared (FIR) to the near-ultraviolet. A Fourier transform spectrometer was
used from 10 cm$^{-1}$ to 18,000 cm$^{-1}$, and a grating monochramotor was
used from 12,000 cm$^{-1}$ to 48,000 cm$^{-1}$. After the reflectance was
measured at various temperatures, the sample was coated in-situ with gold
or aluminum and the measurements were repeated at all temperatures,
providing an absolute measure of the reflectivity.\cite{Homes93} The error
in the absolute value of the reflectance is below 1$\%$. The relative error
in the reflectance measured at different temperatures does not exceed 0.1$\%$.

To obtain the complex optical constants the Kramers-Kronig relations were
used. In order to extend the reflectance data to higher energies the
reflectance of Pr$_{1.85}$Ce$_{.15}$CuO$_4$ from 6 eV to 38 eV was adopted.
\cite{Arima93} Above 38 eV the functional form R $\propto\omega^{-4}$ was
assumed. When extrapolating the reflectance to zero frequency a Hagen-Rubens
model was applied in the normal state, and a two-fluid model in the
superconducting state. The error in the reflectance has been propagated to
the optical constants and combined with uncertainties introduced by the
extrapolation procedure, and will be discussed in the text.

\section{Reflectivity measurements and Kramers-Kronig analysis}

\subsection{a-axis}

Fig.~\ref{fig:Rab} shows the a-axis reflectance of the SC (Panel A) and AF
(Panel B) samples in the FIR. The 292 K spectrum of the SC sample shows
metallic behavior with a single phonon at 300 cm$^{-1}$. There is additional
weak structure between 400 cm$^{-1}$ and 700 cm$^{-1}$ which grows in
intensity as the sample is cooled. The reflectance of the SC sample shows a
strong temperature dependence in the FIR. By 25 K the reflectance is nearly
5$\%$ above that at room temperature. Between 150 cm$^{-1}$ and 300 cm$^{-
1}$ the reflectance of the SC sample at 25 K rivals that of such excellent
conductors as Cu, Al, and Au yet the DC transport measurements indicate that
the sample is a rather poor conductor.\cite{Hidaka89,Andrade93,Jiang94} The
resolution of this ambiguity can be found in the reflectance data below 150
cm$^{-1}$. Instead of R($\omega$) approaching 1 monotonically as
$\omega\rightarrow0$ as in a metal, the reflectance decreases giving rise to
a peak in R($\omega$). As will be demonstrated later, this is a consequence
of having poor conductivity at $\omega$=0 which greatly increases at finite
frequencies. One final peculiarity of the SC sample is the lack of
temperature dependence below T$_c$. The superconducting (7 K) reflectance
increased slightly below 100 cm$^{-1}$, but at all higher frequencies the
spectra was identical to the data at T$_c$ (25 K) within 0.1$\%$. The
fact that the sample is truly a bulk superconductor can be confirmed by
Fig.~\ref{fig:Rc} were the c-axis reflectance changes dramatically below
T$_c$.

The FIR reflectance of the AF sample (Panel B) is qualitatively different
from it's SC counterpart. The reflectance is lower at all temperatures and
drops more quickly with increasing frequency. Notice that the y-axis of the
top panel covers only 8$\%$ while the bottom panel extends from 85$\%$ -
100$\%$. The phonon mode seen at 300 cm$^{-1}$ is still clearly visible
here. Three other modes that have been previously reported\cite{Onose99} can
also be identified at 130, 340, and 510 cm$^{-1}$. In addition a broad
"hump" structure extending from 250 - 450 cm$^{-1}$ is observed which grows
in intensity as the temperature is lowered. No downturn in
R($\omega\rightarrow0$) is observed and at low temperatures R($\omega<$ 300
cm$^{-1}$) is nearly independent of frequency.

The right panel of Fig.~\ref{fig:Rab} (C) shows the a-axis reflectance over
an extended frequency range for both the SC and AF samples at 292 K and 25
K. The dominant feature of $R(\omega)$ in both samples is a plasma minimum
at $\omega\sim$ 11,000 cm$^{-1}$. The reflectance of the SC sample is
$\sim$ 10$\%$ higher than that of the AF
sample in the mid-infrared, but drops below it near the plasma minima.
In the SC sample R($\omega$) at 25 K smoothly
joins the room temperature reflectance before the plasma minimum. In
contrast to this behavior, in the AF sample the 25 K  R($\omega$) curve
crosses the 292 K spectrum at 1,000 cm$^{-1}$ and in the low temperature
spectrum reveals a partial gap-like depression at $\omega$ $<$ 4,000 cm$^{-1}$.
This structure has previously been reported by Onose and coworkers.
\cite{Onose99}

Fig.~\ref{fig:sigmaSC} shows the real part of the conductivity,
$\sigma_1$($\omega$), for the a-axis of the SC sample generated from the
R($\omega$) data in Fig.~\ref{fig:Rab}. The most prominent feature is the
peak in $\sigma_1$($\omega$) below 100 cm$^{-1}$. This should be contrasted
with the behavior of conventional metallic systems which can be described by
the Drude model:
\begin{equation}
\label{eq:drude}
\sigma_1^D(\omega) = \frac{\omega_p^2\tau}{1+\omega^2\tau^2}.
\end{equation}
where $\tau$ is the charge carrier lifetime, and $\omega_p$ is the plasma
frequency which is determined by the ratio of the carrier density to effective
mass. This form of the conductivity is a Lorentzian centered at $\omega=0$.
The peak in $\sigma_1(\omega)$ at $\omega\neq0$ observed in NCCO signals a
departure from conventional metallic transport. This is also evident from
the non-monotonic behavior of R($\omega$) displayed in the top panel of
Fig.~\ref{fig:Rab}. The peak in $\sigma_1(\omega)$ grows in strength and
softens as the temperature is lowered. In the superconducting state the peak
frequency shifts to 55 cm$^{-1}$ and peak height is slightly reduced. The
inset of Fig.~\ref{fig:sigmaSC} shows the 292 K and 25 K spectra, where the
y-axis is nearly an order of magnitude smaller than in the main figure, and
the scale of the x-axis extends up to 4,000 cm$^{-1}$ ($\simeq$ .5 eV). It
is clear that the peak is a robust feature seen even at room temperature.
Also notice that the 25 K curve is non-monotonic above the peak,
showing a local minimum  at 400 cm$^{-1}$. At higher frequencies
$\sigma_1(\omega)$ decays slower than expected from Eq.~\ref{eq:drude}.

In Fig.~\ref{fig:sigmaAF} $\sigma_1$($\omega$) is shown for the AF sample.
The spectrum of $\sigma_1$($\omega$) at 292 K is flat and featureless. At 25
K a broad resonance is observed at 50 cm$^{-1}$ $<$ $\omega$ $<$ 300
cm$^{-1}$. The AF sample has a
semiconducting resistivity, so as $\omega\rightarrow0$ the 25 K curve is
expected to drop below the 292 K spectrum. As in the SC sample a minimum can
be seen near 400 cm$^{-1}$ at 25 K. The inset shows $\sigma_1$($\omega$)
over an extended energy region for several different temperatures. At low
temperatures an additional channel of absorption develops giving rise to a
maximum at $\omega$ $\simeq$ 2,200 cm$^{-1}$.\cite{Onose99}

\subsection{c-axis}

The c-axis reflectance for both the SC (solid lines) and
AF (dashed-doted line) samples is shown in Fig.~\ref{fig:Rc}.
The two samples are very similar over most of the
frequency range. The FIR reflectance is dominated by three strong
phonon modes at 134 cm$^{-1}$, 268 cm$^{-1}$, and 507 cm$^{-1}$ (right panel).
At higher energies a flat featureless reflectance is observed up to 40,000
cm$^{-1}$ ($\sim$ 5 eV),
after which the reflectance begins to rise. Both samples showed some
temperature dependence in the phonon region. Above 700 cm$^{-1}$
the reflectance of both samples was independent of temperature. Also note
that at $\omega$ $>$ 60 cm$^{-1}$ the reflectivity of the SC sample is the same above
and below T$_c$. The left panel shows the detailed temperature dependence
of the SC sample in the low energy region. At T = T$_c$ the reflectance
is nearly flat with just a slight upturn as
$\omega\rightarrow$ 0. As the sample becomes superconducting a plasma
minimum develops due to the screening by the superconducting current.
This minimum deepens and moves to higher frequencies as the temperature
is reduced.

Some of the key parameters defining the c-axis electrodynamics of the normal
and superconducting state can be
determined directly from the raw R($\omega$) data. For this purpose we show
the low frequency R($\omega$) data for the SC sample at 7 K and 25 K in the
left panel of Fig.~\ref{fig:sigmaC}. The location of the minimum in
R($\omega$) is the screened plasma frequency, $\omega_p^*$. The screened
plasma frequency quantifies the superfluid density, $\rho_s$, $\omega_p^*$ =
$\sqrt{\rho_s/\epsilon_{\infty}}$, where $\rho_s\propto$ n$_s$/m$^*$ and
n$_s$ is the density of paired electrons and m$^*$ is their effective mass.
The reflectance can be expressed in terms of the dielectric function as:
R($\omega$) = $(\frac{\sqrt{\epsilon}-1}{\sqrt{\epsilon}+1})^2$. By letting
$\epsilon$ = {\it constant}, we can account for the nearly flat
reflectance at T$_c$ and just above the plasma minimum at lower temperatures
below the frequency of the first phonon. Specifically a value of
$\epsilon$ = 17, gives a constant reflectance of 37$\%$ (dashed
line). With this value of $\epsilon$ and the location of the
minimum in R($\omega$) we determine $\rho_s$ =  3600 cm$^{-2}$ which
corresponds to a c-axis penetration depth of $\lambda_c$ = c/$\sqrt{\rho_s}$
= 26 $\mu$m. This value is confirmed by an analysis of the dielectric
function generated from a Kramers-Kronig transformation of R($\omega$).
Along these same lines we can gain an estimate for the DC conductivity at
T$_c$ directly from R($\omega$). We model the upturn seen in the reflectance
of the 25 K spectrum as $\omega\rightarrow$ 0 with $\epsilon$ and
Eq.~\ref{eq:drude}. A reasonable assumption for the spectra seen in
Fig.~\ref{fig:sigmaC} is 1/$\tau\gg$ 100 cm$^{-1}$, therefore there is only
one free parameter, $\omega_p^2\tau$ =  $\sigma_{DC}$, and we obtain an
excellent fit to the data with $\sigma_{DC}$ = 1.5 $\Omega^{-1}$cm$^{-1}$
(dotted line).

The right panel of Fig.~\ref{fig:sigmaC} shows
$\sigma_1$($\omega$) obtained from Kramers-Kronig transformation of
R($\omega$) for the SC sample. The frequency range is confined to the
region below the first phonon mode ($\omega<120$ cm$^{-1}$). The
electronic contribution to the conductivity is extremely weak. For example,
$\sigma_1$($\omega$ = 10 cm$^{-1}$) = 1.5 $\Omega^{-1}$cm$^{-1}$ at 25 K,
which is the same value as was obtained from the fit of R($\omega$) in the
left panel. Below T$_c$ the temperature dependence of the low frequency
conductivity is anomalous; at 10 cm$^{-1}$ $\sigma_1$(T$_c$) $>$
$\sigma_1$(T = 7 K), however by 50 cm$^{-1}$ $\sigma_1$(T$_c$) has
dropped below the conductivity at 7 K. Also notice that $\sigma_1$(T = 19 K)
is {\it greater than} $\sigma_1$(T$_c$) throughout the entire frequency
range depicted in Fig.~\ref{fig:sigmaC}.

An example of the error in $\sigma_1$($\omega$) is shown at 95 cm$^{-1}$ for
the T = 25 K spectrum. The error was calculated by propagating the uncertainty
in the reflectance, and taking into account variations caused by different
extrapolations of the reflectance to high and low frequencies. For the
present analysis it is important to distinguish between absolute  and
relative errors. The absolute error is shown by the large bars to be
$\pm$ 1 $\Omega^{-1}$cm$^{-1}$. The relative error in the temperature
dependence is an order of magnitude smaller and shown by the small bars.

\section{Discussion}

\subsection{Electron dynamics in the CuO$_2$ plane: Pseudogap}

A unique characteristic of the electron doped cuprates is the manner in
which superconductivity is induced in the phase diagram. The as-grown
crystals of Nd$_{2-x}$Ce$_x$CuO$_{4+\delta}$ progresses from an insulator at
x=0 to a metal at x=.21 without the appearance of a superconducting phase.
The superconducting state can only be realized by annealing the as-grown
crystals in an oxygen free atmosphere.\cite{Takagi89} While this procedure
reduces the oxygen content by only $\simeq$ 1$\%$\cite{Schultz96} the
changes in $\sigma_1$($\omega$) through out the infrared are significant.
What can be drawn from Fig.~\ref{fig:sigmaSC} and Fig.~\ref{fig:sigmaAF} is
that the AF sample is under doped with respect to  the SC sample. To
quantitatively compare the differences in $\sigma_1$($\omega$)
Fig.~\ref{fig:SW} shows $\sigma_1^{SC}(\omega)-\sigma_1^{AF}(\omega)$ below
14,000 cm$^{-1}$. In the SC sample the mid-infrared absorption is reduced,
while the low frequency ($\omega < 1,300$ cm$^{-1}$) absorption increases.
The inset shows the effective spectral weight, N$_{eff}(\omega)$ =
$\int_{0}^{\omega}\sigma_1$($\omega'$)d$\omega'$, for both the SC and AF
sample. While the low energy spectral weight grows more quickly in the SC
sample, N$_{eff}(\omega)$ is nearly equal by 12,000 cm$^{-1}$ ($\simeq$ 1.5
eV) for the two materials.\cite{Neff} This effect is similar to the result of
Ce doping from x=.12 to x=.2 in oxygen reduced Pr$_{2-x}$Ce$_{x}$CuO$_{4+\delta}$.
\cite{Cooper90} The deoxyengenation process that takes the AF sample into the
superconducting phase is similar to the doping processes in hole doped
cuprates. Spectral weight is transferred to lower energies which enhances the
metallic response and induces superconductivity.

We now turn to the analysis of the evolution of electron dynamics associated
with changes of carrier density from the under doped (AF) region to the
optimally doped (SC) sample. A useful optical constant within the context of
this discussion is the frequency dependent scattering rate:\cite{Puchkov96}
 \begin{equation}
\label{eq:tau}
1/\tau(\omega)={\omega_{p}^2 \over {4\pi}} Re({1 \over \sigma(\omega)}).
 \end{equation}
In Fig.~\ref{fig:tau} we plot 1/$\tau$($\omega$) for both the AF sample
(left panel) and the SC sample (right panel). Looking first at the right
panel we see that above $\omega$ $\sim$ 650 cm$^{-1}$ 1/$\tau$($\omega$)
varies nearly linearly with $\omega$. However, at lower frequencies
1/$\tau$($\omega$) drops faster than
this linear trend. The low frequency suppression is strongest at 25 K, yet
persists even at room temperature. The top of the "shoulder" in
1/$\tau$($\omega$) is chosen as the frequency, $\Theta$,
characterizing the low energy depression of 1/$\tau$($\omega$).

Turning to the AF sample we again see the low frequency suppression of
1/$\tau$($\omega$) which is now much more pronounced. In addition, the
magnitude of the scattering rate is nearly twice that of the SC sample.
However, the characteristic frequency remains at $\Theta$ = 650 cm$^{-1}$ as
in the SC sample. The frequency dependence of 1/$\tau$($\omega$) also
remains nearly linear above $\Theta$,\cite{lowT} but there is less
temperature dependence in this region than found in the SC sample.

Several features of 1/$\tau$($\omega$) described above for NCCO are
characteristic of the pseudogap state in hole doped cuprates.
\cite{Timusk99,Puchkov96,Puchkov96i,Basov96}
Fig.~\ref{fig:tauBi} shows typical 1/$\tau$($\omega$) spectrum for under and
weakly over hole doped Bi2212.\cite{Puchkov96i} Examining first the under
doped compound in the left panel, the most prominent feature is the
depression of 1/$\tau$($\omega$) below $\simeq$ 700 cm$^{-1}$ in both the
normal and superconducting state. This depression is absent at 292 K were
the linear high-$\omega$ trend continues to the lowest frequencies, yet is
clearly well developed at $T>T_c$. This reduction
of the scattering rate has been attributed to the opening of a partial gap,
or pseudogap, in the density of states. In Bi2212 and other hole doped
cuprates the pseudogap has been shown to have the same $d_{x^2-y^2}$
symmetry and similar magnitude as the superconducting gap.\cite{Shen}
Turning to the over doped sample in the right panel we see that the feature
in 1/$\tau$($\omega$) is now only observed in the superconducting state. The
overall magnitude of the scattering is much smaller, and at high frequencies
1/$\tau$($\omega$) is temperature dependent. The doping dependent trends of
1/$\tau$($\omega$) are consistent with a variety of other
measurements:\cite{Timusk99} in the under doped phase a pseudogap opens at a
temperature T$^*$ $>$ T$_c$, with an energy scale $\Theta$ $\simeq$
2$\Delta$, while the pseudogap is absent above T$_c$ in the over doped
phase.

Looking back at the NCCO data in Fig.~\ref{fig:tau} we can see several
features that are common to 1/$\tau$($\omega$) in the hole doped cuprates.
The onset of the depression in 1/$\tau$($\omega$) remains at
$\Theta$ = 650 cm$^{-1}$, independent of doping. The magnitude
of 1/$\tau$($\omega$) is higher in the under doped samples, and the onset of
the depression is much sharper. As doping increases the low energy
suppression of 1/$\tau$($\omega$) decreases. In addition a strong temperature
dependence is observed across the energy range displayed for the more
heavily doped samples. There are a few
marked differences with the NCCO data. First, 1/$\tau$($\omega$) in the
superconducting state (not shown) is nearly identical to the 25 K spectrum above
150 cm$^{-1}$. This is a result of an anomalously small superfluid density,
as will be discussed in the next section. The second main difference is that
the depression in 1/$\tau$($\omega$) is observed at all temperatures,
suggesting T$^*$ exceeds 292 K in both the AF and SC samples. This latter
result may resolve a long standing discrepancy between the linear temperature
dependent resistivity of optimally doped hole cuprates\cite{Takagi92}
and the nearly quadratic temperature dependent resistivity in NCCO. In the
pseudogap state ($T<T^*$) $\rho(T)$ decreases faster than linear with
decreasing temperature, possibly accounting for the
anomalous behavior found in NCCO. In fact recent experiments have found a
nearly linear $\rho(T)$ at T $>$ 292 K in NCCO.\cite{Greene00}

The observation of a pseudogap in NCCO (T$_c$ = 25 K) with a characteristic
temperature T$^*$ $>$ 300 K, has significant implications on our theoretical
understanding of this phenomenon. In particular, our data suggest
that the pseudogap may be distinct from the superconducting gap. Determining
the relationship between
the pseudogap and superconducting state has become a central problem in the
field of high temperature superconductivity.\cite{Buchanan01} Because many
electronic and magnetic properties such as Fig.~\ref{fig:tauBi} (left panel) and
others\cite{Shen} evolve smoothly from the pseudogap to superconducting
state, the pseudogap is often considered a precursor of the superconducting
gap. In addition, the similarities of the energy scales associated with the
pseudogap and superconducting states has also been used to support this
view. In contrast, in the present experimental work these energy scales are
much different. The pseudogap energy scale, $\Theta$, extracted from
1/$\tau$($\omega$) is roughly the same as found in the hole doped cuprates
(about 5 - 10$\%$ less). However, recent photoemission work\cite{Armitage00} on NCCO
indicate that the superconducting gap is as small as 2$\Delta$ $\simeq$ 3
meV, more than an order of magnitude smaller than $\Theta$. These data
clearly suggest that in NCCO the pseudogap, as determined from
1/$\tau$($\omega$), is not the same as the superconducting gap. Whether this
conclusion will hold in other cuprate were the two energy scales are very
similar will require further study.

A complimentary interpretation of the pseudogap structure in 1/$\tau$($\omega$)
may be formulated in terms of charge carriers coupling to a collective mode.
\cite{Shen97,Shen01}
At the energy of the mode a new channel of scattering opens for the charge
carriers leading to an increase in 1/$\tau$($\omega$).\cite{Allen71} From an
inversion of the 1/$\tau$($\omega$) curve the spectrum of the collective mode
$W(\omega)$ can be estimated;\cite{Carbotte99}
\begin{equation}
\label{eq:mode}
W(\omega) = {1 \over {2\pi}}{d^2 \over {d\omega^2}}\left( \omega {1 \over {\tau(\omega)}} \right).
\end{equation}
The top panel
of Fig.~\ref{fig:mode} shows 1/$\tau$($\omega$) at T$_c$ along with the above inversion.
The spectra of $W(\omega)$ shows a clear peak at $\omega$ = 400 cm$^{-1}$.
Also included in Fig.~\ref{fig:mode} are a representative collection of 1/$\tau$($\omega$)
spectra and their corresponding $W(\omega)$ for several
different families of electron and hole doped cuprates including single,
double, and triple layer materials.\cite{Basov96,Puchkov95,Startseva99,McGuire99}
In all of the materials inversion of
1/$\tau$($\omega$) produces similar structure in $W(\omega)$.
Recently $W(\omega)$ derived from the 1/$\tau$($\omega$) spectra in YBCO has
been attributed to antiferromagnetic fluctuations seen as a 41 meV peak in
inelastic neutron scattering experiments.\cite{Carbotte99} However, this peak
has only been observed in YBCO\cite{Fong96,Dai96} and Bi2212\cite{Fong99}.
Clearly, the remarkable
similarities of 1/$\tau$($\omega$) and therefore $W(\omega)$ found
in different families of cuprates shown in
Fig.~\ref{fig:mode} calls for a uniform description of this feature, rather
than ad hoc scenarios for each compound.

While the microscopic origin of the low $\omega$ depression in
1/$\tau$($\omega$) is unresolved, it is clearly a feature common to all
families of cuprate superconductors. This global similarity of the low
energy charge dynamics is in agreement with the suggestion that the
pseudogap is a generic feature of the doped Mott-Hubbard
insulator.\cite{Huscroft01} In addition these results support the idea
that the key features of the low energy charge dynamics in high-T$_c$
cuprates are the same for both electron and hole doped materials.

\subsection{Localization in the Cuprates}

We now turn to the discussion of the low frequency in-plane conductivity.
As pointed
out earlier the SC sample shows a peak in $\sigma_1(\omega)$ below 100
cm$^{-1}$. The peak is unusual because it signals a qualitative departure
from a free carrier response. In a wide variety of elemental metals the
response of the charge carriers is adequately represented by the Drude model
(Eq.~\ref{eq:drude}).\cite{Ordal83} More complex materials in which
$\sigma_1(\omega)$ deviates from the Lorentzian form suggested by
Eq.~\ref{eq:drude} the conductivity still remains monotonic as
$\omega\rightarrow0$. The peak
in $\sigma_1$($\omega$) at a finite frequency is indicative of charge
carrier localization.\cite{loc} In the regime of low dimensionality the
electron gas is
particularly susceptible to localization effects arising from disorder.
Examples can be found in 1-D organic conductors\cite{Kohlman} and 2-D field
effect transistors\cite{Allen75}. Localization effects should also not be
surprising in the cuprates where the nearly decoupled CuO$_2$ layers give
rise to a quasi 2-D system. In optical studies of controlled
induced disorder in ion irradiated YBa$_2$Cu$_3$O$_{6.95}$\cite{Basov94} and
Zn doped YBa$_2$Cu$_4$O$_8$\cite{Basov98} a finite frequency peak in
$\sigma_1$($\omega$) progressively developed with the introduction of
disorder. Other cuprates may show "intrinsic" disorder due to the
substitutional process which controls the charge carrier density and induces
superconductivity. NCCO may fit in this latter category,\cite{inhomo}
because the oxygen reduction process leaves a random distribution of oxygen
vacancies. Another example of an "intrinsically" disordered system is
Bi$_2$Sr$_{2-x}$La$_x$CuO$_4$\cite{Weber00} where an inhomogeneous
distribution of La and Sr doping may give rise to the observed localization
effects.\cite{peak}

To compare with the NCCO results Fig.~\ref{fig:sigmaBi} shows an example of
peaks in $\sigma_1(\omega)$ in hole doped Bi$_2$Sr$_{2-x}$La$_x$CuO$_4$ for
two different levels of La doping.\cite{Weber00} The main panels show the
conductivity in a similar fashion to Fig.~\ref{fig:sigmaSC} and the inset'
show the R($\omega$) data. Three temperatures are displayed; T = 10 K, T$_c$
(28 K for x=.3 and 25 K for x=.4), and 292 K. The most important feature for
this discussion is the obvious maximum seen in R($\omega$) spectra near 150
cm$^{-1}$ in both crystals. As discussed earlier the non-monotonic behavior
of R($\omega\rightarrow$ 0) is a clear indication of a finite frequency peak
in $\sigma_1$($\omega$). As the main panels show the peaks in
$\sigma_1$($\omega$) are similar to that seen in NCCO
(Fig.~\ref{fig:sigmaSC}). The peak in the x=.4 sample is at 90 cm$^{-1}$ and
in the x=.3 sample it is slightly lower but less well defined. In the
superconducting state the peak in both samples softens.

One puzzle with the localization effects observed in the optical
conductivity of NCCO and other cuprates is the persistence of "metallic"  DC
resistivity (positive $d\rho/dT$). Conventionally the peak in
$\sigma_1(\omega)$ is accompanied by an activated DC transport. Moreover,
the frequency  of the peak in $\sigma_1(\omega)$ agrees well with the
activation energy extracted from the resistivity data for 2-D electron gas
in Si field-effect transistors.\cite{Allen75} Nevertheless, a coexistence of
the "metallic" resistivity and the peak in $\sigma_1(\omega)$ is a robust
result reported for a variety of cuprates. This enigma can be qualitatively
understood by considering the complexity of the Fermi surface in the
cuprates.\cite{Shen93} The DC conductivity primarily probes the
quasiparticles at the zone diagonal where the $d$-wave gap has nodes. The
optical conductivity is also dominated by the nodal regions. However,
electronic states at the zone boundaries [$(\pi,0)$ or $(0,\pi)$] are also
sampled through the IR measurements as evidenced through the observation of
the $d$-wave pseudogap.\cite{Puchkov96,Basov96} According to recent
photoemission results, with the systematic addition of impurities, states
develop within the gap at ($\pi$,0).\cite{Vobornik99} This has recently been
confirmed through the observation of intra-gap resonance's with scanning
tunneling microscopy.\cite{Pan} We believe that these resonance's may
be connected with the peak observed in $\sigma_1(\omega)$. At the same time the DC
transport is effectively shunted by the highly mobile nodal quasiparticles
and therefore remains relatively insensitive to the dramatic changes close
to the zone boundary.

The effects of disorder are also expected to be prominent in the
superconducting state. In a $d$-wave superconductor disorder leads to pair
breaking,\cite{Radtke93} and will therefore decrease the total amount of
superfluid.
Infrared studies have shown that with increasing disorder the superfluid
density ($\rho_s$) is systematically depleted while a concomitant growth of
the finite frequency peak in $\sigma_1(\omega)$ is observed.
\cite{Basov94,Basov98} In NCCO we also observe an anomalously small
superfluid density. This can be demonstrated from the following sum
rule:\cite{sumrule}

\begin{equation}
\label{eq:rho}
\rho_s = \frac{120}{\pi}\int_{0^+}^{W_c}[\sigma_1^N(\omega) - \sigma_1^S(\omega)]d\omega
 \end{equation}
where the superscripts N and S refer to the normal and superconducting state.
Applying Eq.~\ref{eq:rho} to the data in Fig.~\ref{fig:sigmaSC} at 7 K and
25 K with the integration cut-off W$_c$ $>$ 1 eV we obtain 4*10$^7$ cm$^{-2}$.
This is only 4$\%$\cite{error} of the spectral weight
of 10$^8$ cm$^{-2}$ which corresponds to the penetration depth of
$\sim$1500$\AA$ reported by microwave and magnetic measurements.
\cite{Wu93,Gollnik98,Nugroho99} A similar suppression of $\rho_s$ is found
in the data for Bi$_2$Sr$_{2-x}$La$_x$CuO$_4$ in Fig.~\ref{fig:sigmaBi} and
was also reported for other cuprates where a peak in $\sigma_1(\omega)$ is
observed.\cite{Basov94,Basov98,Weber00,Startseva99,McGuire99}

In contrast to our results, Homes et al.\cite{Homes97} found a value of
$\lambda$ = 1600$\AA$ from the above sum rule analysis, and did not observe
a finite frequency peak in $\sigma_1(\omega)$. One possible explanation is
that the T$_c$ = 23 K crystal was slightly over doped with respect to the
crystals measured in this study. In the over doped phase the interlayer
coupling of the CuO$_2$ planes increases and a more 3-D character develops,
making  samples less susceptible to localization effects. Doping dependent
studies on NCCO and other cuprates showing signs of localization would help
clarify this matter.

\subsection{Interlayer Transport}

In contrast to the changes in the optical response of the a-axis when
varying the oxygen content, there is very little difference between the SC
and AF samples along the c-axis (Fig.~\ref{fig:Rc}). The matrix element that
dominates interlayer transport in cuprates has been show to have the same symmetry as
the $d$-wave gap: nodes at ($\pi$,0) and anti-nodes at
($\pi$,$\pi$).\cite{Xiang98} Therefore our polarized infrared measurements
can be viewed as an indirect probe of doping dependent changes in the Fermi
surface topology. If this interpretation is applied to the results shown in
Fig.~\ref{fig:Rc}, one can conclude that the oxygen reduction procedure
that induces superconductivity has very little effect on the Fermi surface
near ($\pi$,0). This is also confirmed by the observation of a pseudogap
(which is a ($\pi$,0) effect\cite{Shen}) in the
in-plane scattering rate for both the AF and SC samples. In contrast,
Figures 1-3 show that the low energy
excitations on other parts of the Fermi surface are dramatically altered.
The oxygen reduction process appears to have the largest impact near the
zone diagonals. Angle resolved photoemission and other k-dependent probes
would be useful in confirming this result.

The most obvious change in the interlayer transport with oxygen doping is
the low frequency plasmon that develops due to the presence of
superconducting carriers. From the location of this plasmon a value of
$\lambda_c$ = 26 $\mu$m is found (Section III-B), in good agreements with
Terahertz transmission measurements on thin films of NCCO.\cite{Pimenov00}
It has been shown\cite{Basov94i,sasa00} that a universal correlation exists
between $\lambda_c$ and $\sigma_1$($\omega\rightarrow0$, T$_c$) in the hole
doped cuprates. This correlation is given by $\lambda_c^{-2}$ =
$\Omega\sigma_1$($\omega\rightarrow0$, T$_c$), where $\Omega$ is related to
the energy scale from which the superfluid is collected. The  dark solid
line in Fig.~\ref{fig:Bplot} represents this correlation for hole doped
cuprates.\cite{sasa00} Other anisotropic superconductors that have
conventional normal and superconducting properties also follow this
correlation (thin line) but have a smaller constant of proportionality
$\Omega$. Organic superconductors, Josephson junctions prepared from elemental
metals, transition metal dichalcogenides, and granular films belong to this latter
group. Our results for NCCO are shown in the plot as a star. NCCO clearly
belongs to the same universality class as the hole doped cuprates. The
origin for the different scaling behavior between the cuprates and
other anisotropic superconductors is as yet unresolved, but may be related
to the incoherent nature of the conductivity discussed below.

In order to understand the impact of the superconducting transition on the
interlayer transport it is useful to examine the distinct contributions to
$\sigma(\omega)$. In the superconducting state the conductivity can be
represented by two components:
\begin{equation}
\label{eq:2fluid}
\sigma^s(\omega) =\sigma^{pair}(\omega) + \sigma^{reg}(\omega).
\end{equation}
The first term represents the paired
carriers and is given by $\sigma_1^{pair}$($\omega$) =
$\rho_s\delta$($\omega$=0)/8. The second term corresponds to unpaired
carriers below T$_c$ and is plotted in the right panel of
Fig.~\ref{fig:sigmaC}. While $\sigma_1^{pair}$($\omega$) is outside of the
range of our experiment, the formation of $\rho_s$ which gives rise to this
term can be seen in the energy loss function, Im(-1/$\epsilon$), plotted in
the inset of Fig.~\ref{fig:sigmaC}. In a conducting material the loss
function shows a peak at a frequency proportional to the carriers plasma
frequency. Looking at the plot of Im(-1/$\epsilon$) we see that at T = T$_c$
(thin black line) the loss function is flat and featureless corresponding to
an over-damped plasmon. As the temperature is reduced below T$_c$ a sharp
resonance develops signaling the formation of $\rho_s$.

The development of $\rho_s$ must be contrasted with the behavior of
$\sigma_1$($\omega$) in Fig.~\ref{fig:sigmaC}. From the sum rule in
Eq.~\ref{eq:rho} we see that as $\rho_s$ increases there should be a
corresponding decreases in $\sigma_1$($\omega$) at finite frequencies. What
is actually observed is much different. At 19 K, where the peak in the loss
function indicates a non-zero superfluid density, $\sigma_1$($\omega$) is
larger than the normal state curve at all frequencies shown in
Fig.~\ref{fig:sigmaC}. In fact above 50 cm$^{-1}$ the absorption is greater
at all temperatures in the superconducting state compared to the normal
state curve at 25 K. In order to better elucidate this behavior we rewrite
Eq.~\ref{eq:rho} as $\rho_s$ =
$\frac{120}{\pi}\int_{0^+}^{W_c}$[$\sigma_1^N(\omega)$ -
$\sigma_1^S(\omega)$]d$\omega$ + $\Delta$K, where we have split the integral
in two\cite{Hirsch}:
\begin{equation}
\Delta K =
\frac{120}{\pi}\int_{W_c}^{\infty}[\sigma_1^N(\omega) - \sigma_1^S(\omega)]d\omega.
\label{eq:DK}
\end{equation}
$\Delta$K represents the contribution to $\delta$(0) from the experimentally
inaccessible integration region. For the c-axis, above 120 cm$^{-1}$
the electronic component of the conductivity is overwhelmed by the response
of the phonons. Therefore the limit of
integration is taken as W$_c$ = 120 cm$^{-1}$, which corresponds to
$\sim$ 8$\Delta$.\cite{Armitage00}.
While the absolute value of W$_c$ is small, it is worth
noting that in all hole doped cuprates that have been studied,\cite{Basov99,Basov01}
$\sigma_1^S\simeq\sigma_1^N$ all ready at 2$\Delta$, and this equality
continues throughout the experimentally available range ($\sim$ 20$\Delta$).

The analysis of Eq.~\ref{eq:DK} reveals the energy scale of the electronic
states that make up the superconducting condensate. In order to determine
the relative amount of $\rho_s$ originating from $\omega$ $>$ W$_c$
we define the normalized missing spectral weight (NMSW) as
$\Delta$K(T)/$\rho_s$(T). This ratio gives the fraction of the total $\delta$-function
spectral weight drawn from $\omega>$W$_c$=8$\Delta$. The NMSW
is plotted as a function of reduced temperature in Fig.~\ref{fig:DK}. This
fraction is close to 1 below $.5T_c$, but increases well above 1 as
$T\rightarrow T_c$. The inequality $\Delta$K(T)/$\rho_s$(T)$>$1 means that
in addition to the growth of $\rho_s$ at $\omega$=0, the finite frequency
spectral weight at $\omega$ $<$ W$_c$ {\it increases} below T$_c$. While
there has been an effort to understand the NMSW in other cuprates where
$0<\Delta$K$/\rho_s<1$,
\cite{Hirsch,Anderson,Chakravarty,Ioffe,Carbotte00,Imada} there is currently
no theoretical basis in which to understand the inequality
$\Delta$K/$\rho_s$ $>$ 1.

In order to clarify the role of the cut-off $W_c$ in the analysis of the
energy scales involved in the formation of the superconducting condensate
we repeat our analysis with a model BCS system.
In conventional superconductors which follow the BCS formalism the formation
of $\rho_s$ is always
accompanied by a {\it decrease} in low frequency spectral weight at
T$<$T$_c$, in accord with Eq.~\ref{eq:rho}. The inset of Fig.~\ref{fig:DK}
gives an example of $\sigma_1(\omega)$ for a BCS superconductor in the dirty
limit\cite{Zimmermann91}. In the
superconducting state of the model $\sigma_1(\omega)$ is zero below $2\Delta$, then
increases sharply and merges with the normal state conductivity. The effect
of W$_c$ on Eq.~\ref{eq:DK}  for this model calculation is shown in the main
panel. For all values of W$_c$ the NMSW is constant with temperature.
The NMSW is always less than 1, regardless of the choice of
W$_c$. For W$_c$ = 2$\Delta$, half of $\rho_s$ is drawn from $\omega>W_c$.
However, with W$_c$ = 8$\Delta$, similar to our experimental cutoff, only
7$\%$ of $\rho_s$ comes from $\omega>W_c$. This example merely reflects
the fact that in the BCS model of superconductivity the electronic states
that form the condensate lie near the Fermi energy. In contrast, the
differences seen in the NMSW of NCCO indicate that $\rho_s$ is being
collected from an extended energy range, by far exceeding the
superconducting gap.

In BCS theory, there is only one energy scale involved in superconductivity:
the superconducting gap, $\Delta$. However, the data for NCCO suggests that
$\Delta$ plays little if any role in determining the region from which
$\rho_s$ is collected. This indicates that there is an additional energy
scale which is greater than 8$\Delta$, in cuprate
superconductors.\cite{Hirsch,Imada}

Studies on the doping dependence of the NMSW in
Tl$_2$Ba$_2$CuO$_{6+\delta}$\cite{Katz00} and
YBa$_2$Cu$_3$O$_{6+\delta}$\cite{Basov01} have found this effect to be
largest in under doped compounds. The conclusion drawn from these experiments
is that when the normal state is incoherent, with low spectral weight, the
superfluid is derived from high energies. As a more coherent response
develops at low frequencies, the superfluid is formed from this spectral
weight near $\omega$ = 0. The incoherent conductivity which drives the source of
$\rho_s$ to higher energies has been linked to the normal state pseudogap.
\cite{Basov01} For technical reasons we were not able to observe a pseudogap
in the c-axis conductivity of NCCO.\cite{pgap}  However, the extremely small
values of $\sigma_1(\omega)$ along with the absence of any obvious Drude peak
in the interlayer conductivity is consistent with strongly incoherent
transport. The source of the large NMSW shown in Fig.~\ref{fig:DK} is likely
to be tied to the incoherent conductivity as in the hole doped cuprates.

\section{Conclusions}

To conclude we find several similarities between NCCO and the hole doped
cuprates. The analysis of 1/$\tau$($\omega$) in the SC sample provides strong
evidence for a pseudogap in the electron side of the cuprate phase
diagram. In addition, the trends seen in the evolution of 1/$\tau$($\omega$)
from the under doped AF sample to the optimally doped SC sample closely follow
the behavior of the hole doped cuprates. In NCCO the pseudogap energy scale
$\Theta$ is more than an order of magnitude greater than 2$\Delta$. This
result implies that, at least in NCCO, the pseudogap and superconducting
gap are not the same.

The peak seen in the in-plane conductivity, along with a low value
of $\rho_s$ is often observed in disordered hole doped cuprates. Disorder
in low dimensional materials, such as the cuprates, often results in charge
carrier localization. However, the non-trivial topology of the Fermi surface
in the cuprates may lead to a coexistence of localization features in
$\sigma_1(\omega)$ with metallic DC transport.

A sum rule analysis of the c-axis conductivity reveals similar trends
as is observed in hole doped cuprates. Namely, the spectral weight that
is transferred to the delta-function peak at $\omega$ = 0 below T$_c$
originates from an energy range in excess of 8$\Delta$. This is fundamentally
different from the behavior of
conventional superconductors, and indicates a large energy scale is involved
in superconductivity in both electron and hole doped cuprates. In addition a
comparison of the normal and superconducting properties in NCCO clearly
places it in the same universality class as the hole doped cuprates as
opposed to other conventional layered superconductors.

This work was supported by the NSF grant DMR-9875980 and the Alfred P.~Sloan
Foundation. D.N.~Basov is a Cotrell Fellow of the Research Corporation.

$^*$Present address: Kohzu Seiki Co.,  Setagaya 1-8-19, Tokyo, Japan

\begin{figure}
\caption{Panel A: a-axis far-infrared reflectance of the SC sample at T =
292, 225, 150, 80, 25, and 7 K. As the temperature drops a well defined
maximum develops at 150 cm$^{-1}$. The temperature dependence below T$_c$ is
confined to $\omega<100$ cm$^{-1}$. Panel B: a-axis reflectance of the AF
sample at the above temperatures. Notice that the y-axis covers a broader
range than in panel A. The overall reflectance is lower than that of the SC
sample and additional phonon modes are visible. An anomalous structure
between 250 cm$^{-1}$ and 450 cm$^{-1}$ is visible at 292 K and grows in
strength as the temperature is reduced. Panel C: high frequency reflectance
of the SC and AF sample at 292 K and 25 K. The screened plasma frequency
(minimum in R($\omega$)) is the same in both samples. While the 25 K and 292
K spectrum merge smoothly in the SC sample, the 25 K curve in the AF sample
crosses below 292 K at 1,000 cm$^{-1}$ and remains suppressed up to 4,000
cm$^{-1}$.} \label{fig:Rab} \end{figure}

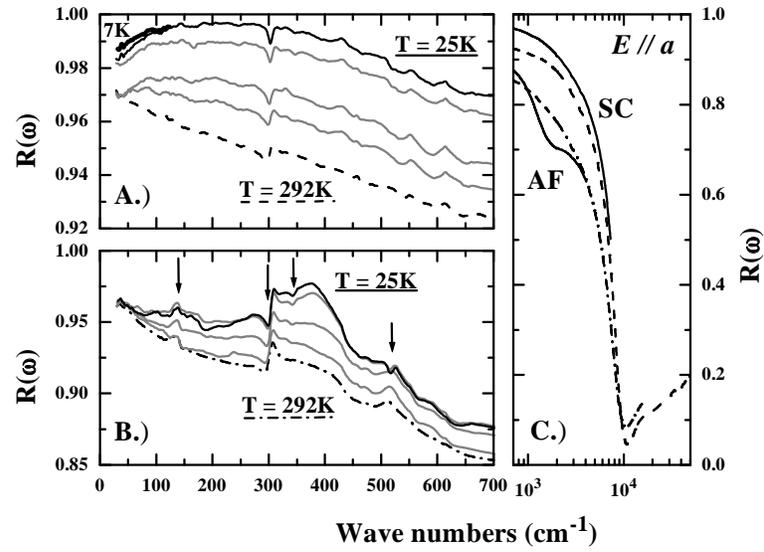
\begin{figure}
\caption{Far-infrared conductivity for the a-axis of the SC sample.
At all temperatures there is a finite frequency peak in $\sigma_1$($\omega$).
The peak grows in intensity and softens as the temperature is lowered to
T$_c$. Below T$_c$ the peaks intensity is slightly reduced as it further
softens. The inset shows $\sigma_1$($\omega$) at 25 K and 292 K up to 4,000
cm$^{-1}$ where the y-axis has been reduced by nearly an order of magnitude.
The peak is clearly visible even at 292 K. The inset shows the canonical
pseudogap behavior of $\sigma_1$($\omega$) as seen in the under doped hole
cuprates. As the temperature decreases the low frequency peak narrows
below a characteristic frequency ($\simeq$ 650 cm$^{-1}$).}
\label{fig:sigmaSC}
\end{figure}

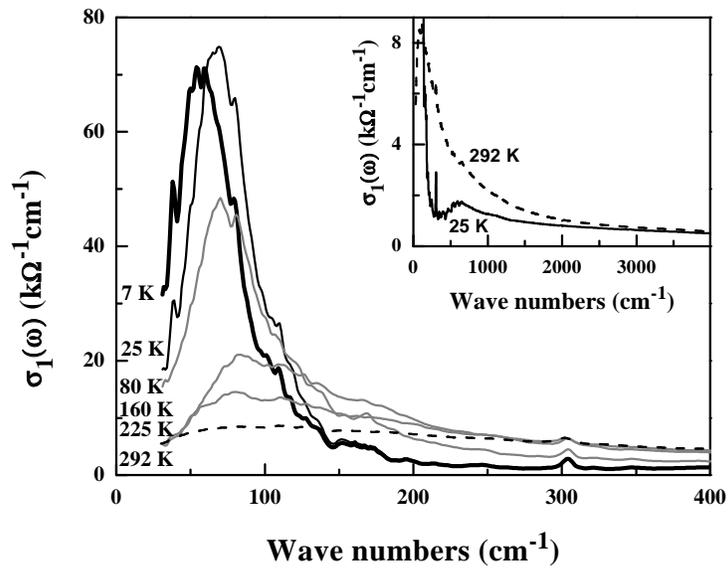
\begin{figure}
\caption{A-axis conductivity of the AF sample at 25 K and
292 K. While $\sigma_1$($\omega$) is flat and featureless at 292K, a broad
resonance develops at 25 K. In addition the 25 K spectrum shows a minimum
near 400 cm$^{-1}$ similar to the SC sample. The inset shows a
second channel of absorption at $\omega$ $\simeq$ 2,200 cm $^{-1}$ that
develops at low temperatures.}
\label{fig:sigmaAF}
\end{figure}

\begin{figure}
\caption{(Right Panel) c-axis reflectance at 25 K of both the SC (solid
lines) and AF sample (dashed-doted line). Above 100 cm$^{-1}$ the spectra of
both samples are nearly identical. Three strong phonons are observed in the
far-infrared. The left panel shows the sub-Terahertz reflectance at 25 K and
lower temperatures. A
slight upturn is seen in the normal state spectra as $\omega\rightarrow0$
indicating a finite electronic contribution to the conductivity. As the
sample becomes superconducting the characteristic plasma edge develops due
to the screening of the superconducting carriers. (Temperatures shown: T =
25, 21, 19, 17, 15, 12, and 7 K)} \label{fig:Rc} \end{figure}

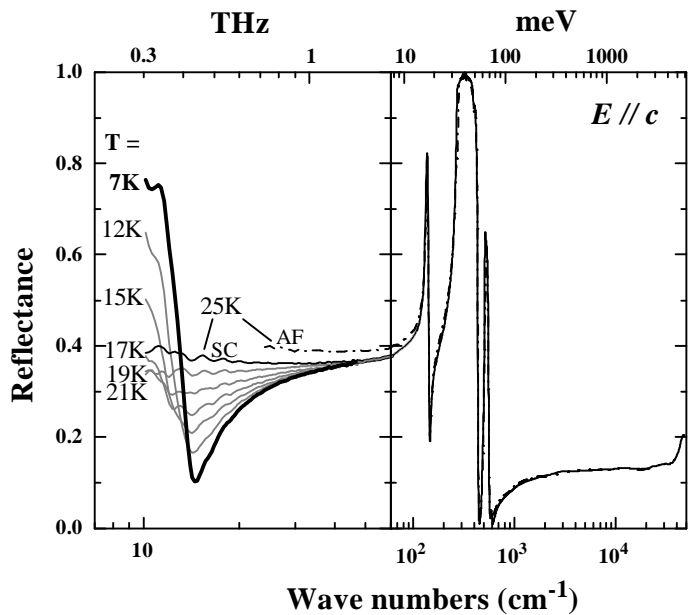
\begin{figure}
\caption{Left Panel: low frequency R($\omega$) at 7 K and 25 K reploted on a
linear scale. The straight dashed line corresponds to R($\omega$) calculated
with only the contribution of $\epsilon$ = 17. From the location
of the minimum in R($\omega$) and this value of $\epsilon$ we
find $\lambda_c$ = 26 $\mu$m as described in the text. Additionally
using $\epsilon$ and Eq.~\ref{eq:drude} we fit
(dashed line) R($\omega$) at 25 K with $\sigma$($\omega=0$) as the free
variable  and obtain a best fit with $\sigma$($\omega=0$) = 1.5
$\Omega^{-1}$cm$^{-1}$.
Right Panel:
electronic contribution to $\sigma_1$($\omega$) in the SC sample below
the first phonon. Notice the non-trivial temperature and frequency
dependence above and below T$_c$. The inset shows the peak in the
loss function due to the development of the superconducting condensate.}
\label{fig:sigmaC}
\end{figure}

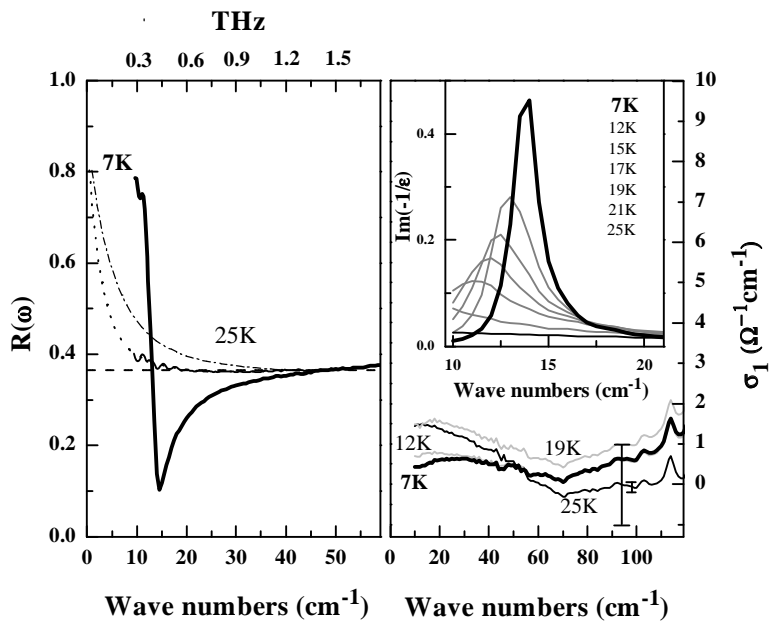
\begin{figure}
\caption{The a-axis differential conductivity,
$\sigma_1^{SC}(\omega)-\sigma_1^{AF}(\omega)$ plotted throughout the infrared.
The oxygen reducing procedure leads to spectral weight transfer from
the mid-infrared
to lower energies. The inset shows that the total spectral weight below
$\simeq$ 14,000 cm$^{-1}$.}
\label{fig:SW}
\end{figure}

\begin{figure}
\caption{The in-plane scattering rate (Eq.~\ref{eq:tau}) for the SC (right
panel) and AF (left panel) samples. Above 650 cm$^{-1}$ in the SC sample the
frequency dependence is linear. At $\omega$ $<$ $\Theta$
1/$\tau$($\omega$) is suppressed faster than a linear extrapolation of the
high frequency data. This suppression can be seen at all temperatures, but
is most pronounced at T$_c$. In the superconducting state (not shown) the
spectrum is nearly the same as at T = T$_c$. The AF sample shows a similar,
but sharper threshold (arrow) also at 650 cm$^{-1}$. These spectra should
be compared to the 1/$\tau(\omega)$ data for hole doped materials in
Fig.~\ref{fig:tauBi}}
\label{fig:tau}
\end{figure}

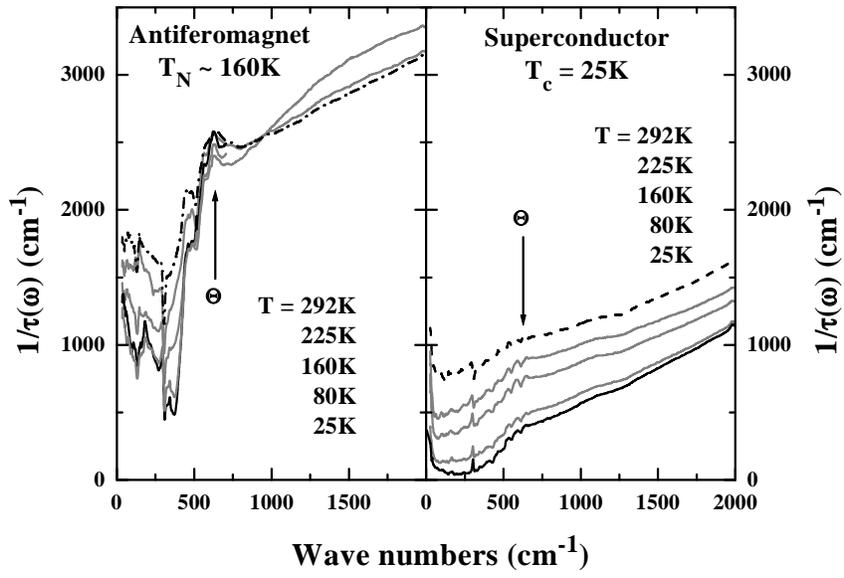
\begin{figure}
\caption{The in-plane scattering rate is plotted at 292 K, T$_c$, and 10 K
for under doped Bi2212 (left panel) and slightly over doped Bi/Pb2212 (right
panel)\cite{Puchkov96i}. In the under doped compound a gap in
1/$\tau$($\omega$) (marked by arrows) opens well above T$_c$, while in the
over doped compound the gap only opens in the superconducting state. Above
the gap 1/$\tau$($\omega$) is independent of temperature for the under doped
system, while it is temperature dependent at all frequencies in the
over doped phase.}
\label{fig:tauBi}
\end{figure}

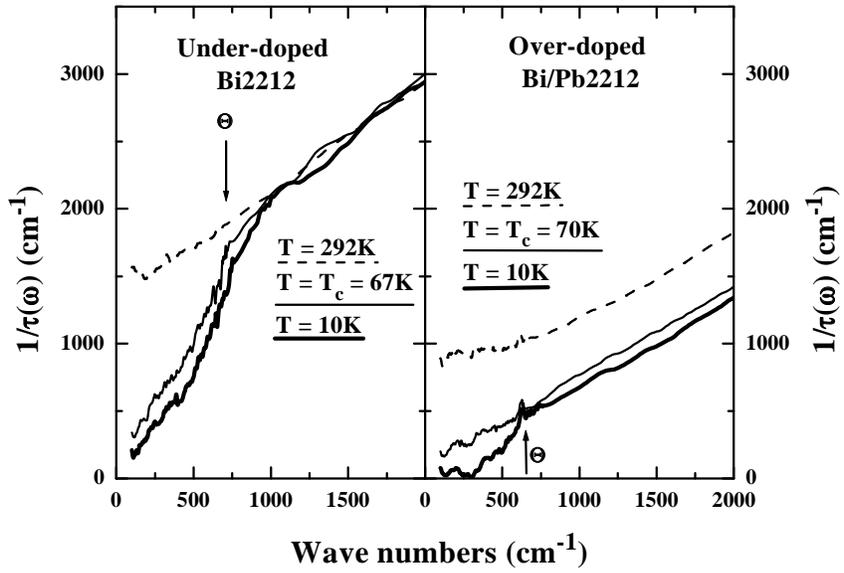
\begin{figure}
\caption{Left Panel: 1/$\tau$($\omega$) at T $\simeq$ 10 K for
several different families of high-T$_c$
cuprates.\cite{Basov96,Puchkov95,Startseva99,McGuire99} All the
spectra share a similar low-$\omega$ depression with a nearly
linear frequency dependence at higher energies. Right Panel:
$W(\omega)$ derived from 1/$\tau$($\omega$) using
Eq.~\ref{eq:mode}. Again, the spectra of $W(\omega)$ all have
similar form.} \label{fig:mode}
\end{figure}

\begin{figure}
\caption{Main Panels: $\sigma_1$($\omega$) within the CuO$_2$ plane for
Bi$_{2-x}$La$_x$Sr$_2$CuO$_4$ with x = .3 (top panel) and x = .4
(bottom panel).\cite{Weber00} A finite frequency peak
can be seen below 100 cm$^{-1}$ in both samples for all temperatures (300 K,
T$_c$, and 10 K shown). The inset show the reflectance data where clear
maximum can be seen near 150 cm$^{-1}$ for both dopings.}
\label{fig:sigmaBi}
\end{figure}

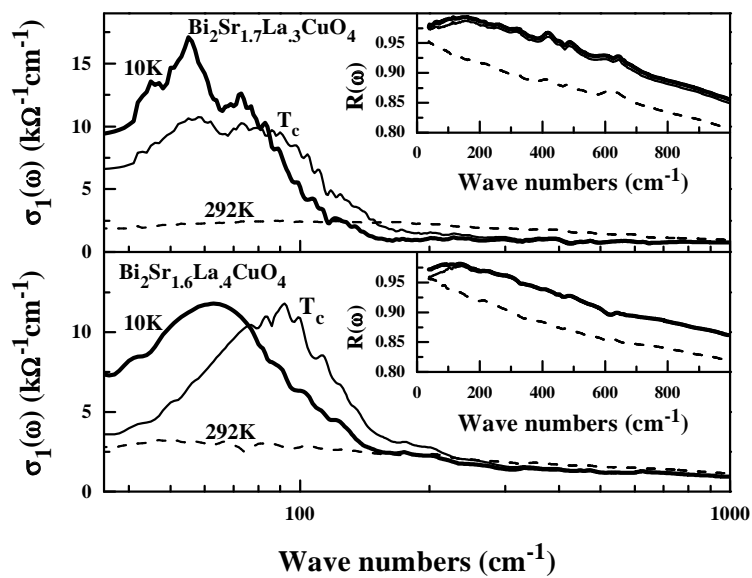
\begin{figure}
\caption{Universal plot showing the correlation between $\sigma_{DC}$(T=T$_c$)
and $\lambda_c^{-2}$ in layered superconductors.\cite{Basov94i,sasa00}
The bottom line corresponds
to the hole doped cuprates. The top line includes 2D organic superconductors,
transition metal dichalcogenides, granular metal films, and Josephson
junctions prepared from elemental metals. Our measurements indicate that
NCCO (star in plot)
belongs to the same universality class as the hole doped cuprates.}
\label{fig:Bplot}
\end{figure}

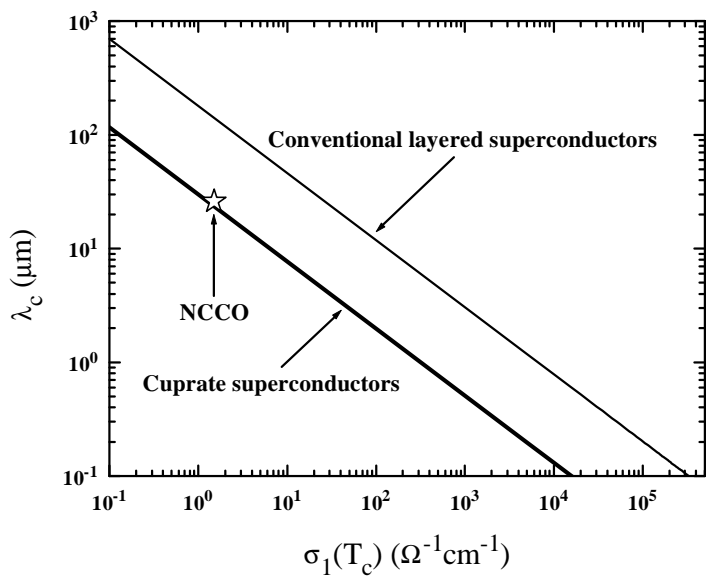
\begin{figure}
\caption{The normalized missing spectral weight (defined in text),
$\Delta$K(T)/$\rho_s$(T),
show the fraction of the superfluid density collected from $\omega>8\Delta$
as a function of reduced temperature. The inset shows a calculation of
$\sigma_1(\omega)$ above and below T$_c$ for a BCS superconductor in the
dirty limit. The lines at the bottom of the main panel shows how
$\Delta$K(T)/$\rho_s$(T) in this model system depends on the integration
limit (W$_c$) in
Eq.~\ref{eq:DK}. Notice that with W$_c$ = 8$\Delta$ only 7$\%$ of $\rho_s$
is drawn from higher energies.}
\label{fig:DK}
\end{figure}


\begin{references}
\bibitem{hpgap} A gap-like feature in $\sigma_1(\omega)$ has been observed in
NCCO by Onose and coworkers at .3 eV.\cite{Onose99} Similar structure has
not been detected in other cuprates. The typical energy scale of the
pseudogap in the hole doped cuprates is $\sim$ 90 meV.
\bibitem{Neff} The different carrier densities for the two samples
produced by the oxygen reducing procedure would dictate that N$_{eff}$ is
crossing at 14,000 cm$^{-1}$ rather than saturating to a constant value.
\bibitem{lowT} For $\omega$ $>$ 700 cm$^{-1}$
1/$\tau$($\omega$) is not shown for the AF sample at 25 K and 80 K, as another channel of
absorption opens and the scattering rate description is not appropriate at
these energies.
\bibitem{inhomo} An alternative explanation for these results is that the
systems are macroscopically inhomogeneous, consisting of metallic and
insulating regions. The mixed phase system can give rise to a peak in
$\sigma_1(\omega)$.\cite{Kohlman}
\bibitem{error} It should be noted that there are substantial errors in the
absolute value of $\sigma_1$($\omega$) near the peak. The absorption is
proportional to 1-R($\omega$), which from examination of Fig.~\ref{fig:Rab}
is very small at low temperatures. Therefore our error in the absolute
values of R($\omega$) of $\pm$ 1 $\%$ causes a large uncertainty in
$\sigma_1$($\omega$). For example, a reduction of the 25 K reflectance by 1
$\%$ cuts the magnitude of the peak in $\sigma_1$($\omega$) by 50 $\%$.
However, the relative error which is an order of magnitude smaller is the
relevant error when applying Eq.~\ref{eq:rho}. Taking this error into
consideration Eq.~\ref{eq:rho} still accounts for less than 5$\%$ of the
spectral weight corresponding to published values of the penetration depth.
\bibitem{pgap} The observation of a pseudogap requires the analysis of the
frequency dependence of $\sigma_1$($\omega$) throughout the FIR. Over
most of this range the electronic contribution to $\sigma_1$($\omega$) is
overwhelmed by the phonon response, therefore completely masking any changes
in $\sigma_1$($\omega$) from the electronic channel.
\end{references}
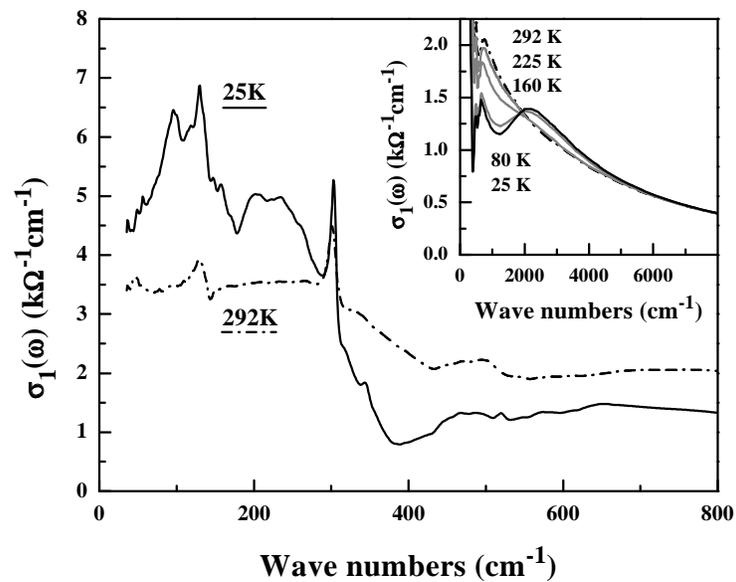
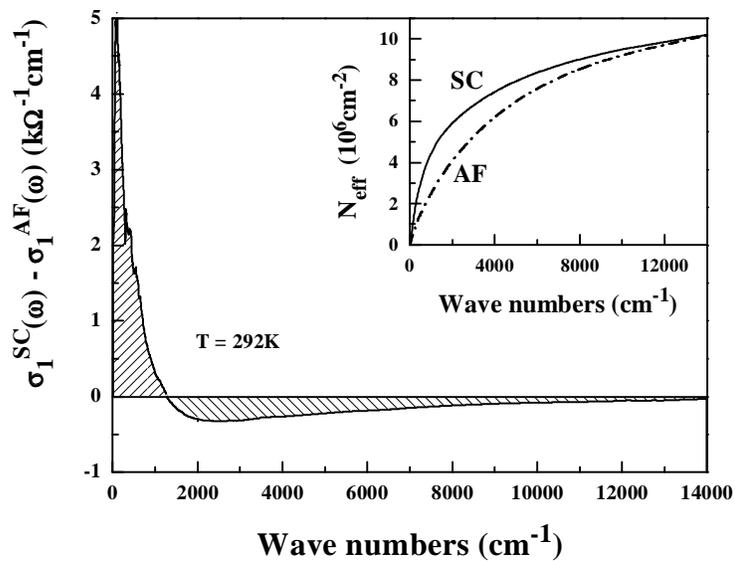
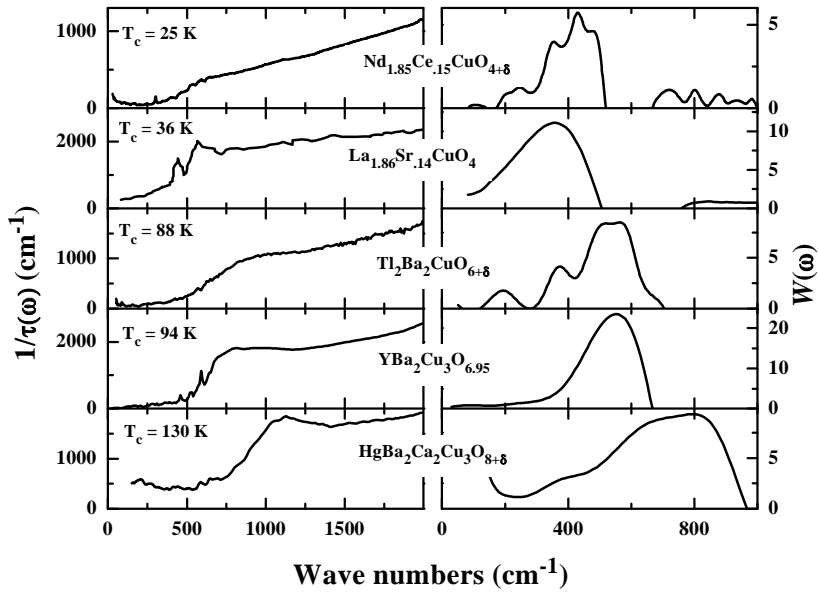
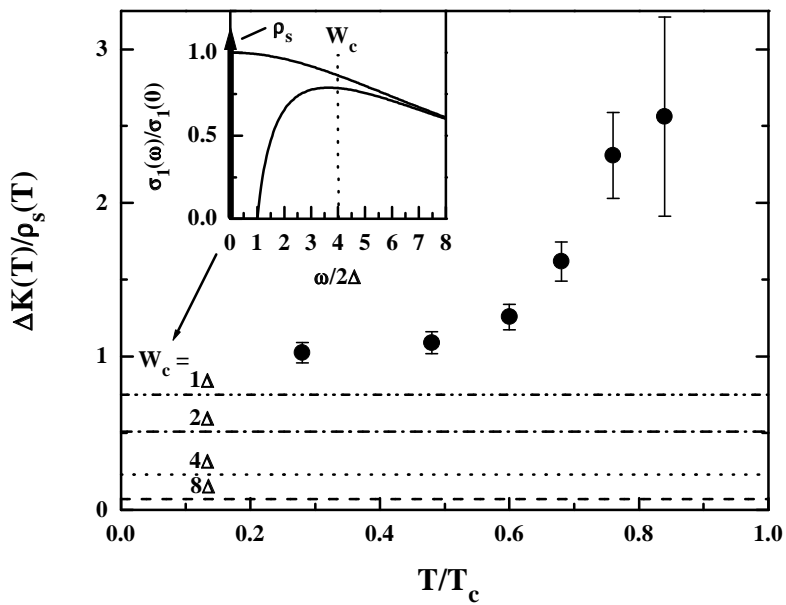
\end{document}